# Nucleosome Chiral Transition under Positive Torsional Stress in Single Chromatin Fibers

Aurélien BANCAUD[1,$,£], Gaudeline Wagner[1,£], Natalia Conde e Silva[2], Christophe Lavelle[2,3,4], Hua Wong[3], Julien Mozziconacci[3], Maria Barbi[3], Andrei Sivolob[5], Eric Le Cam[4], Liliane Mouawad[6], Jean-Louis Viovy[1], Jean-Marc Victor[3*] and Ariel Prunell[2*]

[1] Institut Curie (CNRS-UMR 168), 75231 Paris, France
[$] Present address : LAAS-CNRS, 7 avenue du Colonel Roche 31077 Toulouse, France

[2] Institut Jacques Monod (CNRS-UMR 7592), 2 place Jussieu, 75251 Paris Cedex 05, France

[3] Laboratoire de Physique Théorique de la Matière Condensée (CNRS-UMR 7600), 4 place Jussieu, 75252 Paris Cedex 05, France

[4] Laboratoire de Microscopie Moléculaire et Cellulaire (CNRS-UMR 8126), Institut Gustave Roussy, 39 rue Camille Desmoulins, 94805 Villejuif, France

[5] Department of General and Molecular Genetics, Taras Shevchenko National University, 64 Vladimirskaya street, 01033 Kiev, Ukraine

[6] Integrative Imaging Unit (INSERM U759), Institut Curie-Recherche, Centre Universitaire Paris-Sud, Bâtiment 112, 91405 Orsay, France

Running title: Nucleosome chiral transition upon positive torsion

£ These authors, in alphabetical order, contributed equally to the work

*Correspondence to  prunell@ccr.jussieu.fr, or victor@lptl.jussieu.fr

**Summary**
	Using magnetic tweezers to investigate the mechanical response of single chromatin fibers, we show that fibers submitted to large positive torsion transiently trap positive turns, at a rate of one turn per nucleosome. A comparison with the response of fibers of tetrasomes (the $(H3-H4)_2$ tetramer bound with ~50 bp of DNA) obtained by depletion of H2A-H2B dimers, suggests that the trapping reflects a nucleosome chiral transition to a metastable form built on the previously documented right-handed tetrasome. In view of its low energy, <8 kT, we propose this transition is physiologically relevant and serves to break the docking of the dimers on the tetramer which in the absence of other factors exerts a strong block against elongation of transcription by the main RNA polymerase.

**Introduction**

The genome of higher eukaryotes is folded into chromatin, whose repetitive motif - the nucleosome - is a nucleoprotein complex with ~160 bp of DNA wrapped around a $(H2A,H2B,H3,H4)_2$ histone octamer (van Holde, 1988). Chromatin structure is dynamic, and this property is essential for DNA transactions such as transcription, which correlates with decompaction of the fiber and alterations at the nucleosome level. In particular, core particles prepared from transcribed chromatin display a deficit of H2A-H2B dimers (Baer and Rhodes, 1983), and a selective dimer exchange with the histone pool was observed in actively transcribed regions (Louters and Chalkley 1985; Schwager et al., 1985 ; Pfaffle et al., 1990 ; Kimura and Cook, 2001). These alterations could be due to the transcription process itself, as observed with RNA polymerase II which displaces *one* dimer *in vitro* (Kireeva et al., 2002), in contrast to prokaryotic RNA polymerases and polymerase III which translocate the octamer (Studitsky et al., 1997). Nucleosome alterations might also be driven by the positive supercoiling generated in front of the polymerase (Liu and Wang, 1987), as directly observed *in situ* (e. g. Ljungman and Hanawalt, 1992). Such an effect of positive supercoiling was previously inferred from *in vivo* experiments with yeast minichromosomes using a conditional topoisomerase mutant to convert DNA supercoiling from negative to positive (Lee and Garrard, 1991). *In vitro*, chromatin reconstitutions on positively supercoiled plasmids showed particles containing their normal histone complement but unable to store negative supercoiling (Jackson, 1993).

Here, we further investigate the influence of positive supercoiling on nucleosome structure through torsional manipulation of single chromatin fibers with magnetic tweezers. After application of a large positive torsion in low-salt, their extension-vs.-torsion response becomes hysteretic, reflecting nucleosomes transition to a transient altered state which traps one positive turn. Based on a comparison with fibers of tetrasomes obtained through depletion of H2A-H2B dimers using three different procedures, we propose that the transition involves i) a breaking of the docking of the dimers on the $(H3-H4)_2$ tetramer ; ii) a switching of that tetramer from its left-handed to its right-handed chiral conformation (Hamiche et al., 1996) ; and iii) an undetermined rearrangement of the dimers insuring that the compaction of the resulting particle is about the same as that of the canonical nucleosome. The real-time transition is slow under low salt conditions, which accounts for the fiber hysteretic behavior, but it gathers pace under more physiological conditions, showing its intrinsic dynamic character. These findings are discussed in the context of *in vivo* transcription.

**Results**
**The standard torsional response**

Nucleosome fibers were reconstituted on a template consisting of 2x18 tandemly repeated 190 or 208 bp 5S nucleosome positioning sequences. Such fibers (Figure 1A) show more or less regularly-spaced nucleosomes with occasional gaps and clusters of two or three close-packed nucleosomes lacking linker DNA (arrowheads). These fibers resemble those observed on the same template with atomic force microscopy, for which a detailed study of nucleosome distribution showed the positioning ability of the sequence together with a similar cooperativity in nucleosome location (Yodh et al., 2002). Typical extension-*vs.*-torsion responses of such fibers in low salt, as obtained using magnetic tweezers (Figure 1B), show a broad apex (Figure 2A; blue), previously accounted for by a molecular model involving a dynamic equilibrium between three nucleosome conformations of distinct topologies, depending on the crossing status of entry/exit DNAs, negative, null or positive (Bancaud et al., 2006). Regularly-spaced nucleosomes can undergo these conformational changes, but close-packed nucleosomes were blocked in the open, uncrossed, conformation. Once all regularly-spaced nucleosomes are in the negative or positive state, further torsions result in quasi-linear regimes in which the fiber compacts rapidly. This shortening is associated with the formation of plectoneme-like structures (Bancaud et al., 2006) that accumulate until the fiber end-to-end distance is close to zero for positive torsions. For negative torsions, in contrast, and as observed with naked DNA (Strick et al., 1996), the DNA spacers (see Figure 1B) and possibly the internucleosomal linkers denature, in which case the fiber fails to compact below a certain length that depends on the applied force. The forward and backward curves, as obtained upon increase or decrease of the torsion, respectively, more or less coincide as long as the positive torsion is not increased much beyond the zero-length limit (Bancaud et al., 2006).

**The hysteretic torsional response**

Beyond this limit, i. e. upon the application of typically +70 turns, the backward curve departs from the forward curve (green and blue, respectively, in Figure 2A), showing an hysteresis. A shift of the green curve toward larger rotations is always observed on the positive side, which is reproducible over many cycles of torsions/detorsions. A shift in the same direction is often observed on the negative side, which varies from one fiber to another. It is sometimes virtually negligible (e. g. fiber 2), but it can also be as large as 60 % the shift on the positive side (e. g. fiber 6). This "negative" shift, in contrast to the "positive" shift, is strongly time-dependent, and it disappears when the backward curve is recorded sufficiently slowly (see fiber 5). Out of 21 fibers, the negative shift is 33 ±21 % (mean ±SD) of the positive one. Importantly, the fiber maximal extension in forward and backward curves remains about the same in all cases. This overall behavior does not depend on the repeat length (208 or 190 bp) of the 5S sequence, nor on the sequence of the DNA template (the poly 601 (200 bp) of Huynh et al., 2005, was tested with similar results; P. Recouvreux and N. C. e S., unpublished results).

We conclude that the backward curve as a whole tends to shift to larger rotations relative to the forward curve. Thus, fibers submitted to large positive stress may trap positive turns. This trapping is stable under positive torsions, but leaks to a variable extent under intermediate negative torsions. Under large negative torsions, all positive turns are released, leading to a reinitialized response that follows the blue curve in the next cycle.

The trapping is unlikely related to long-range nucleosome-nucleosome attraction within the fiber potentially favored by the torsion-induced compaction, since this would cause the fiber to shorten, which is not observed. Positive turns may instead be trapped through a structural alteration of individual nucleosomes. Consistently, stable shifts measured on the positive sides of the curves are directly proportional to the number of regularly-spaced nucleosomes (see Experimental Procedures), with a rate of 1.3 ±0.1 turns per such nucleosome (Figure 2B, right panel). In contrast, the shifts are not correlated with the total number of nucleosomes (left panel). This shows that close-packed nucleosomes, which contribute little to the breadth of the torsional response (see above), do not contribute to the hysteresis either. With $\Delta Lk_p \sim -0.4 \pm 0.1$ (Lk is the DNA linking number) being the topological deformation of positively-crossed nucleosomes in the positive plectonemic region (Bancaud et al., 2006), the topological deformation associated with the altered form relative to the unconstrained DNA is $\Delta Lk_a \sim \Delta Lk_p$ +1.3 ~ +0.9 ±0.2.

**Involvement of H2A-H2B dimers in the transition**

In order to test the altered form for the presence of dimers, we investigated the consequence of their removal on the fiber torsional response. In bulk experiments, NAP-1 (Nucleosome Assembly Protein-1; Ishimi and Kikuchi, 1991) removes the first dimer, much less the second (Kepert et al., 2005), from nucleosomes in the open, uncrossed state (Figure 3A, and legend), and none from nucleosomes in the closed negative state (Conde e Silva et al., 2007). To increase the efficiency of dimer removal in the fiber, we combined NAP-1 with traction, which has been shown alone to have the potential to destabilize dimers (Claudet et al., 2005; Mihardja et al., 2006). After rinsing NAP-1 out, backward and onward curves now coincide over several torsion/detorsion cycles (purple in Figure 3B, panel 2), revealing a fiber lengthened by ~ 420 nm or ~ 19 nm/total nucleosome. Given that nucleosomes are ~2-turn particles with 50 nm of wrapped DNA, this extension suggests that most nucleosomes were unwrapped by ~1 superhelical turn as a consequence of the loss of their dimers.

While we ignored whether all dimers were removed, we wished to test the fiber for a potential release of some tetramers. For this, NAP-1-chaperoned dimers were flowed back into the cell and rinsed out after a 5 min incubation. Remarkably, all features of the starting nucleosome fiber, i. e. its length and hysteresis, were rescued (bold blue and green curves in Figure 3B, panel 3). We conclude that: i) our NAP-1-plus-traction treatment did not remove any significant number of tetramers; ii) the hysteresis strictly depends on the presence of dimers, thus the altered nucleosomes must contain them; and iii) the resulting fiber essentially is a fiber of tetrasomes, and the hysteresis-free purple curve in Figure 4B (panel 2) could be attributed to such a fiber.

To ascertain the torsional behavior of the tetrasome fiber, dimers were removed using two alternative means. Heparin (a strong acidic polyelectrolyte) efficiently removes dimers in bulk experiments (Figure S1A in Supplemental Data available with this article online). After a careful choice of the heparin concentration and the use of nucleosome core particles as a dimers acceptor, the resulting fiber (Figure S1B, panel 3) showed an extension of 24 nm/total nucleosome, very similar to that displayed above in Figure 3B, panel 2. In a third experiment, fibers were washed with 0.7 M NaCl, a concentration sufficient in principle to remove a substantial part of the dimers (Wilhelm et al., 1978; Ruiz-Carillo and Jorcano, 1978). After rinsing the salt out, the torsional response again shows similar breadth and fiber extension as those shown in Figure 3B, panel 2, except for a small residual hysteresis presumably due to incomplete dimers removal (red and purple in Figure S1C, panel 2).

**The tetrasome fiber**

A tetrasome fiber shows: i) a relatively extended structure of maximal length intermediate between that of the nucleosome fiber and that of naked DNA; ii) no hysteresis upon return from high positive torsions; and iii) a center of rotation which approximates that of naked DNA. The first feature is consistent with the smaller wrapping in tetrasomes (~0.7 turn) relative to open-state nucleosomes (~1.4 turns), the second with the strict dimer-requirement of the hysteresis, and the third, with tetrasomes known ability to fluctuate between "pseudo-mirror-symmetrical" left- and right-handed chiral conformations of nearly equal and opposite ΔLk [ΔLk = -0.7 and +0.6 (±0.05) for 5S DNA; Sivolob and Prunell, 2004; see Discussion]. The striking overall similarity of the tetrasome fiber response with the nucleosome fiber backward curve with respect to both their breath and center of rotation (compare purple and green curves in Figure 3B, panel 2; Figure S1B, panel 3; and Figure S1C) suggests that the whole nucleosome could, like the tetrasome, switch from a left- to a right-handed conformation. This switching is opposed by a much higher energetic barrier in the nucleosome, due to the presence of dimers, than in the tetrasome, which accounts for the hysteresis in the first, but not the second, fiber.

**Structural dynamics of the tetrasome**

A number of experimental evidences exist for the tetrasome chiral transition (see Discussion). Here we detail the first Normal Modes analysis (NMA; Wilson et al., 1955; Mouawad and Perahia, 2006) of the tetrasome, to identify structure-based cooperative atomic motions within it. The more atoms involved, the more collective the motion, the lower its frequency, and the more useful that motion could potentially be for the dynamics of the complex. The three lowest-frequency modes essentially correspond to movements of the two H3-H4 dimers relative to one another, the rotation components of which are shown in Figure 4A. The first mode is linked to a wing-like motion of the tetrasome, and the third mode to its lateral opening. The second mode, which describes the rotation of one dimer relative to the other, is directly relevant to the chiral transition. The tetrasome was perturbed along the direction of mode 2 toward positive handedness, and then allowed to relax free of any constraint to a local minimum of energy. Figure 4B, right, shows the DNA superhelix of the resulting tetrasome (red) superimposed on the nucleosomal DNA superhelix (green). Clearly, this local minimum corresponds to a right-handed, although relatively flat, superhelix. Next, the tetrasome was perturbed in the opposite direction, i. e. toward a more left-handed superhelix, and again let relax. No local energy minimum was now observed, and the structure went back to its initial conformation (yellow superhelix in Figure 4B, left panel). We conclude that intrinsic tetrasome structural dynamics encompass its switching to a right-handed conformation.

**Energetics of the transition**

The hysteresis, i. e. the metastability of the altered nucleosome, could be attributed to an energetic barrier against the transition. To test this possibility, we monitored the length of a fiber in the backward curve at constant force (dot 1 in Figure 5A, left panel). The fiber was allowed to relax for ~30 min (green recording in Figure 5A, right panel), closer to the onward curve (dot 2). Thus, the altered structure switches back to the canonical state when given enough time. The proportions of each state can be obtained as a function of time (inset in Figure 5A, right panel; and Figure S2), and used to estimate the energy parameters of the transition (see Experimental Procedures, and Supplemental Data). We obtained an equilibrium energy difference of 10 ±2 kT relative to the ground state of the nucleosome (the open state), and an energy barrier of 30 ±5 kT.

Next, we started from the onward curve, i. e. a fiber containing only canonical, positively-crossed nucleosomes (dot 1 in Figure 5B, left panel). The fiber length, monitored for 30 min, showed no significant increase above the back-ground thermal fluctuations (not shown), as expected from the steady-state equilibrium containing mostly canonical nucleosomes (>90 %, as deduced from the above energy parameters of the transition). This suggests that the equilibrium needs to be shifted toward the altered state in order for the transition to be observed in real time from the onward curve. This was done by raising the force step-wise until a fast extension occurred at 3.4 pN (blue recording in Figure 5B, right panel). The force was then decreased to its initial value, leaving a residual extension as a tribute to the formation of the altered form upon mechanical constraint. We note that the requirement for a high tension to trigger the transition provides an additional argument against the contribution of potential nucleosome-nucleosome attraction forces in the hysteresis.

**Effect of salt on the transition**

As described in our previous study (Bancaud et al., 2006), fibers in higher salt tend to progressively compact from one rotation cycle to the next (blue curves in Figure 6A, right panel). This compaction appears to originate from the occurrence of tails-mediated attractive interactions between nucleosomes (Bertin et al., 2004; Korolev et al., 2006). Interestingly, a force of a few pN applied at the center of rotation break these interactions and rescue the initial rotational response, which is close to that obtained in low salt (Bancaud et al., 2006). At the same time, the hysteresis tends to disappear with salt, to become hardly visible in 25 mM NaCl (Figure 6A).

A fiber in the onward curve (Figure 6B, left panel) was found to relax in ~1 min in 50 mM NaCl (right panel), as compared to ~30 min in low salt (see above). This results in a lower equilibrium energy difference, 6 ±2 kT, and energy barrier, 25 ±5 kT (see details in Supplemental Data). Because the transition rate depends exponentially on the energy barrier, the equilibrium between the two states becomes more dynamic in comparison to the time-scale of data acquisition, and onward and backward curves tend to merge toward a unique curve corresponding to an intermediate equilibrium.

## Discussion
### The tetrasome chiral transition

Experimental evidences for that transition have accumulated since the initial observation that tetramer affinity increases almost equally with negative and positive supercoilings in DNA minicircles. Upon electron microscopic examination, tetrasomes assembled on DNA minicircles of ΔLk = -1 or +1, both ~0.7-turn particles, were undistinguishable from each other. The involvement of the protein in the transition was demonstrated through the oxidation of the two H3 cysteines 110 at the H3/H3 interface, which prevented reconstitution either on topoisomer +1 or –1, depending on the thiol reagent used. Moreover, the formation of a disulfide bridge between these two cysteines did not affect the transition. This latter observation, together with the ability of a steric hindrance at the dyad to block the tetramer left- or right-handed, suggested that the transition occurred through a rotation of the two H3-H4 dimers relative to one another around the disulfide bridge (Hamiche et al., 1996; Alilat et al., 1999). An additional piece of evidence was obtained by neutron scattering, which showed that the pattern of tailless octamers exactly matches that predicted from the core particle crystal structure (Luger et al., 1997), in contrast to the pattern of tailless tetramers which substantially differed (Baneres et al., 2001). The discrepancy appeared to be related to an average flattening of the tetramer protein superhelix as a consequence of its chiral flexibility (J. Parello, personal communication).

Another experimental evidence for the chiral transition is provided by the observation that the center of rotation of the tetrasome fiber, i. e. the rotation at which its length is maximal, approximates that of DNA (Figures 3 and S1), and thus must be associated with particles equally distributed between left- and right-handed conformations. Note that this center of rotation may not be the point at which the tetrasome fiber is relaxed. The tetrasome right-handed conformation being energetically less favorable than the left-handed conformation (by ~2 kT under physiological conditions; Sivolob et al 2000), we anticipate that the torque is positive at the center.

Normal Modes analysis (NMA) has proved useful to identify near-equilibrium functional conformational changes within proteins and protein complexes (see Bahar and Rader, 2005, for a review). NMA was applied here to the tetrasome using the all-atom method (see Experimental Procedures). The three lowest-frequency, i. e. most cooperative, vibrational modes correspond to movements of the whole H3-H4 dimers about each other. Strikingly, the axes of their rotation components (Figure 4A) are approximately orthogonal to each other and all pass through the cysteines 110 (the two green balls in Figure 4A, middle). These features make the cysteines a pivot for H3-H4 dimers relative movements, as originally proposed (see above). Mode 2 appears directly relevant to the transition, while mode 3 mediates tetrasome lateral opening (Figure 4A). A combination of modes 2 and 3 was not explored as an attempt to increase the superhelix right-handedness above that of the mode-2-minimized form (red in Figure 4B). The relevance of mode 3 is suggested by topological data which indicate that the tetrasome was ~20 % more laterally opened in the right- than in the left-handed conformation (Sivolob et al., 2000).

**A nucleosome chiral transition**

The topological deformation achieved by the altered nucleosome form is close to that of the right-handed tetrasome (+0.9 against +0.6; see Results). Moreover, the torsional response of the tetrasome fiber is similar to the backward curve of the nucleosome fiber with respect to both its center of rotation and breadth (see Results). These similarities strongly suggest that the *reverse* transition process is common to both particles. The discrepancy observed in the *direct* transitions (the hysteresis of the nucleosome fiber but not of the tetrasome fiber) may then solely reflect the high energy barrier in nucleosomes (see below), linked to the presence of the H2A-H2B dimers. We thus propose that the core of the altered nucleosome is a right-handed tetrasome.

The fiber similar maximal extensions in onward and backward curves necessarily reflects similar length components of canonical and altered nucleosomes along the direction of the force. In other words, the altered nucleosome must be as compact as the open-state nucleosome which predominates at the center of rotation (Bancaud et al., 2006), i. e. the two particles fold or wrap about the same length of DNA.

In the first step of the transition, dimers are expected to break their docking on the tetramer. Whether they also break their binding sites with the DNA (at superhelix locations (SHL) ±5.5, ±4.5 and ±3.5 ; Luger et al., 1997) is unlikely, especially in view of the fact that they would otherwise diffuse into the cell and be lost. Although these contacts, taken individually, may be relatively weak at physiological ionic strength (Li et al., 2005; Tomschik et al., 2005), their cumulative effects, together with the possible help of the H2B/H4 interface, could be sufficient to hold the dimers in place on the DNA, especially in low salt. Older experiments in which nucleosomes were exposed to salt concentrations <0.2 mM support that possibility. Nucleosomes were indeed found to reproductively elongate with no histones loss, which was interpreted as an unfolding, driven by the repulsion between the DNA gyres, of nucleosomes considered as a sequential arrangement of histones on the DNA (van Holde, 1988 and references therein) (Figure 7, scheme 2). Once the docking is broken, the tetramer may undergo the chiral transition (scheme 3).

The classical equation : $\Delta Lk = \Delta Tw + Wr$, where $Wr$ is the writhe, shows that $\Delta Lk$ may include a twist component ($\Delta Tw$). $\Delta Lk$ = +0.6 of the 5S right-handed tetrasome partitions into $Wr$ = +0.3 and $\Delta Tw$ = +0.3 (Sivolob and Prunell, 2004). Assuming a similar value of $\Delta Tw$ on the altered nucleosome (if H2A-H2B dimers do not contribute), we would get: $Wr$ = +0.6 (+0.9 - 0.3). This writhe is intermediate between that of the above tetrasome and that of a *virtual* right-handed nucleosome mirror image of the open-state nucleosome, +1. This suggests that the altered particle, although it folds a similar length of DNA as the open-state nucleosome (see above), is substantially more open. This might in part be due to the H3 $\alpha$N (and N-terminal tails), which are no longer appropriately located to interact with, and stabilize its entry-exit DNAs (Figure 7).

Given the small $Wr$ value of the altered nucleosome and the ~50 bp of DNA wrapped around the tetrasome, how may the 30-40 extra base pairs fold on each side ? Two possible routes beyond step 4 are illustrated in Figure 7. In model I, entering and exiting dimers-bound DNA duplexes may tend to wind around each other. In model II, they may instead try to continue the right-handed superhelix of the tetrasome, helped by the dimers which would somehow extend the tetramer positive superhelical spool.

While we cannot at the moment go much further into the description of our structure, the question may be asked as to its potential relationship with the *lexosome*, an elusive particle proposed to be a specimen of a transcription-poised nucleosome (e. g. Bazett-Jones et al., 1996 ; but see Protacio and Widom, 1996). A main characteristics of the lexosome, and actually the basis for its purification, is the accessibility of its H3 cysteine 110 thiols. In this respect at least, our altered nucleosome may differ from the lexosome. Indeed, oxydation of these thiols into a disulfide bridge does not interfere with the tetrasome chiral transition (see above), suggesting the thiols are similarly inaccessible in the altered nucleosome. In view of this difference, and in as much as its structure is unique, we propose to call it a reversome (for reverse nucleosome).

**The energetic barrier**

The requirement to break dimers docking on the tetramer is expected to be a major contributor to the energy barrier (25-30 kT; see Results). This view is in keeping with microcaloric studies of octamer assembly in 2 M NaCl that gave an estimate of ~17 kT for the binding energy of one dimer onto the tetramer (Benedict et al., 1984). A mechanical (or elastic) barrier may also exist beyond the point of dimers undocking. Twist may accumulate at the expense of writhe, and be suddenly released, generating an instability similar to that predicted for twisted rods (Neukirch et al., 2002). This writhing instability is expected to be enhanced by the histone-imposed DNA curvature, in conjunction with the extra lateral opening of the structure required at mid-transition to relieve the clash between entry/exit DNA arms (Figure 7, scheme 3, and Sivolob et al., 2000). It is not known to which extent protein-protein interactions could develop in models I or II, and contribute to the low equilibrium energy of the transition.

**Physiological relevance**

RNA polymerases exert a positive torque >1.25 kT/rad, and hence can generate an energy >8 kT over one turn (Harada et al., 2001), sufficient in principle to trigger the transition (~6 kT/turn in 50 mM salt). The existence of an energy barrier against the transition, because it determines its rate, raises the question whether reversomes can be produced at a distance in a time-scale consistent with the polymerase elongation speed. We thus implemented a kinetic model in which a fiber is twisted at constant angular velocity, the torsional constraint being relaxed by the nucleosome-reversome transition in a steady-state manner. Given the speed of RNA polymerase II (~20 nucleotides or 2 turns/sec; Epstein and Nudler, 2003, and references therein), the above energy parameters lead to the effective torque involved, ~1.5 kT/rad (M. B. and J.-M. V., unpublished). This figure is close to the above minimal torque value, suggesting that the transition may indeed propagate ahead of a transcribing polymerase. In this regime, the "reversome wave" is expected to progress much faster than the polymerase (~1.5 (2/1.3) reversomes or ~300 bp/sec, against 20 bp/s, respectively) and, assuming the chromatin is decondensed and nucleosomes can rotate relative to one another, may rapidly reach the end of the transcriptional domain. Beyond this point, further progression of the polymerase may rely on the relaxing activities of the endogenous topoisomerases (Salceda et al., 2006 and references therein).

A nucleosome on a short DNA fragment, in which torsional constraints cannot develop due to free rotation of the ends, presents an almost absolute block to *in vitro* transcription by RNA polymerase II at physiological ionic strength. The block is relieved in higher salt (>0.3 M KCl; Kireeva et al., 2002), which favors dimer loss. Consistently, enzymes such as ACF, or elongation factors such as FACT, which promote removal of a dimer, facilitate transcription elongation (Ito et al., 2000; Reinberg and Sims, 2006). [It is interesting that blocks have also been identified within the tetrasome, which are not relieved by FACT, but this situation seems restricted to nucleosomes on strong positioning sequences such as the 601 (Bondarenko et al., 2006).] Thus, the barrier to transcription is likely to be due to dimers docking. Reversomes, in contrast, owing to their open character (see above), may have their dimers relatively destabilized, suggesting they could behave as torsion-driven "activated" nucleosomes poised for polymerase passage. The chiral-switching ability of the tetramer may then be viewed as the lever used by the main RNA polymerase to break dimers docking *via* the wave of positive supercoiling it pushes in its front. The transition "reversibility" observed in higher salt (Figure 6) would insure that reversomes go back into canonical nucleosomes as soon as the polymerase has traversed them and that a negative constraint develops in its wake. The activity of FACT or other intervening factors to remove dimers may then be dispensable, but nature provides numerous examples of biological redundancy, in which distinct mechanisms were devised to the same end.

**Experimental Procedures**
**Magnetic tweezers, nucleosome arrays and electron microscopy**

Magnetic tweezers manipulation and nucleosome arrays preparation were done as described (Bancaud et al., 2006). Electron microscopic visualization (Figure 1A) was performed as follows: 5 µl of reconstituted nucleosome arrays at 1-5 nM in TE (10 mM Tris-HCl [pH 7.5], 1 mM EDTA) were deposited onto a 600-mesh copper grid covered with a thin carbon film activated by glow-discharge in the presence of pentylamine; grids were washed with aqueous 2% (w/v) uranyl acetate, dried and observed in the annular darkfield mode using a Zeiss 902 electron microscope; and images were captured with a Megaview III CCD camera.

**Regularly-spaced and close-packed nucleosomes**

Regularly-spaced and close-packed (linker-free) nucleosomes assemble in variable proportions from one fiber to the next within the same preparation (Figure 1A). The former nucleosomes fluctuate between the three conformations (negative, open and positive, of respective topology $\Delta Lk \sim$ -1.4, -0.8 and -0.4), while the latter are frozen in the open state. Modeling showed that the proportion of nucleosomes in each state at the apex of the length-vs.-rotation curve of an all-regularly-spaced fiber is respectively 20 %, 65 % and 15 % (Bancaud et al., 2006). This leads to a mean topological deformation of -0.85 turn and to an effective extension of 6.5 nm per nucleosome, against $\sim$-0.5 turn and $\sim$4 nm, respectively, for an all-close-packed fiber (156-bp repeat length). These figures lead to accurate fits of experimental length-vs.-rotation data, via the number of regularly-spaced and close-packed nucleosomes (Bancaud et al., 2006).

These numbers can more simply be estimated from the length of the fiber ($L_{fiber}$) compared to that of the corresponding naked DNA at the same force ($L_{DNA}$). Measurements are made at the centers of rotation, which are shifted relative to one another by $\Delta Lk_{fiber}$. Fibers contain 8680 bp total: 36 208-bp nucleosome positioning sequences and two DNA spacers of $\sim$600 bp (Figure 1A). One can write:

$$\Delta Lk_{fibre} = -0.85 n_{reg} - 0.5(n_{total} - n_{reg}) \qquad (1)$$

$$L_{fibre} = L_{DNA}\frac{1200}{8680} + 6.5 n_{reg} + L_{DNA}\frac{208}{8680}(36 - n_{total})$$
$$+ 4(n_{total} - n_{reg}) + L_{DNA}\frac{50}{8680}(n_{total} - n_{reg} - 1) \qquad (2)$$

with $n_{reg}$ the number of regularly-spaced nucleosomes, $n_{total}$ the number of all nucleosomes, and 50 bp the addition to the naked DNA length made by each close-packed nucleosome. Inversion of these equations gives $n_{total}$ and $n_{reg}$. Errors in $n_{total}$ and $n_{reg}$ are estimated to be ±1 and ±2 nucleosomes, respectively.

## Normal Mode analysis (NMA) of the tetrasome

The tetrasome (~50 bp wrapped around the (H3-H4)$_2$ tetramer) was extracted from the nucleosome crystal structure (1KX5 in the protein data bank). The all-atom force field CHARMM27 (McKerell et al., 1998a) was then applied to it. The energy was minimized using the CHARMM program (McKerell et al., 1998b; Brooks et al., 1983) until it reached a root-mean-square gradient less than $10^{-5}$ kcal/mol.Å, necessary for NMA, then 100 lowest-frequency NM were calculated using the DIMB method (Mouawad and Perahia, 1993 and 2006). The rotation components of the first three were calculated using Hingefind (Wriggers and Schulten, 1997; Figure 4A).

## Kinetic modelling of the transition

We consider a monomolecular reaction scheme for the structural transition with

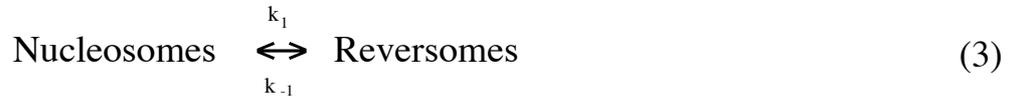

$$\text{Nucleosomes} \underset{k_{-1}}{\overset{k_1}{\leftrightarrow}} \text{Reversomes} \qquad (3)$$

Calling $f$ and $1-f$ the fractions of nucleosomes and reversomes in a fiber, respectively, transition reactions are governed by the following kinetics:

$$\frac{df}{dt} = k_{-1}(1-f) - k_1 f \qquad (4)$$

If the fiber initially contains only reversomes, Equation (4) can be integrated into:

$$f(t) = \frac{k_{-1}}{k_1 + k_{-1}}\left(1 - \exp\left[-(k_1 + k_{-1})t\right]\right) \qquad (5)$$

$k_1$ and $k_{-1}$ can then be fitted and used to derive the energetical parameters of the reaction (see Supplemental Data).

# Figures

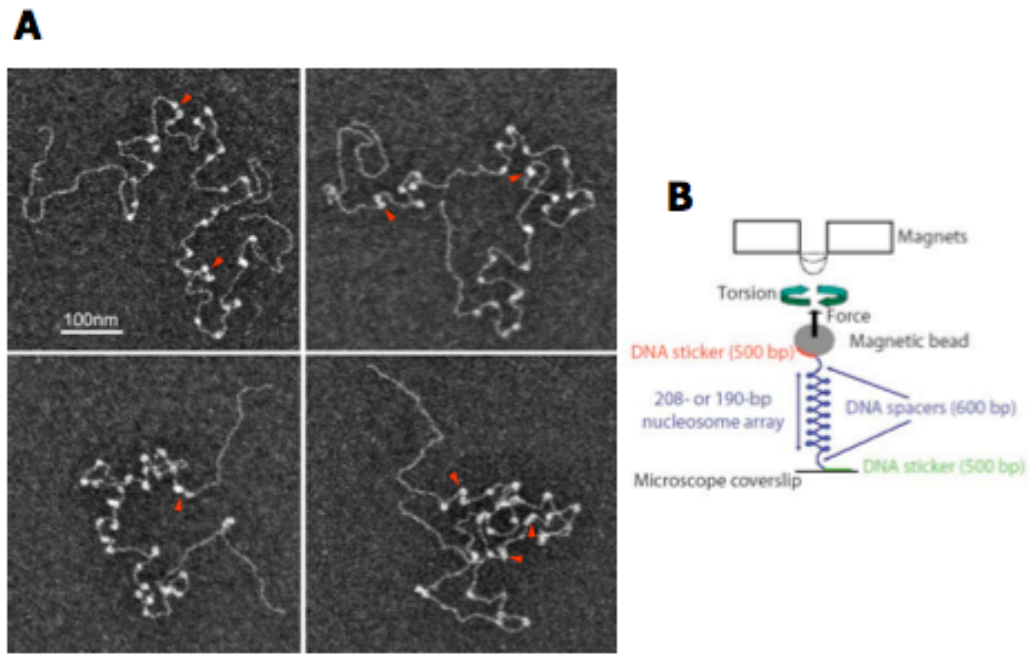

Figure 1. Experimental strategy
(A) Typical "190-bp" fibers before their attachment to the bead and into the flow cell, as visualized by electron microscopy (see Experimental Procedures). Red arrowheads indicate clusters of close-packed nucleosomes. Nucleosome-free DNA spacers and stickers (~1100 bp total; see (B)) flanking the arrays are well visible.
(B) Scheme of the magnetic tweezers set-up (Strick et al., 1996; Bancaud et al., 2006).

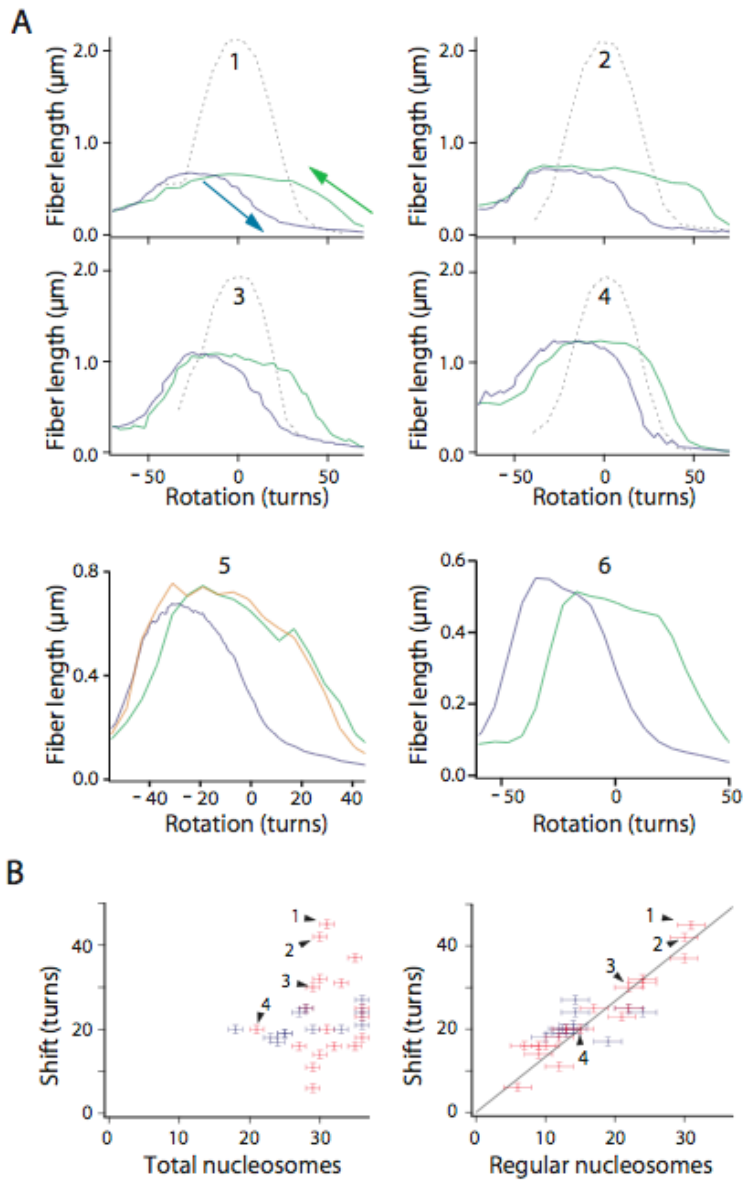

Figure 2. The hysteretic response
(A) Length-*vs.*-rotation responses in $B_0$ (TE plus 0.1 mg/mL BSA) of six "208-bp" fibers at 0.3 ±0.07 pN. The 0-turn rotation reference corresponds to the relaxed state of the corresponding DNAs obtained after histone removal (black dashed curves; Bancaud et al., 2006). After acquisition of the onward curves (blue), the backward curves (green) obtained when returning from high torsions reveal an hysteresis. Independent of the shift observed on the *positive* side, the shift on the *negative* side varies, being minimal in fiber 2 and maximal in fiber 6. Panel 5 shows the dependence of the "negative" shift (but not of the "positive" shift) on the speed of data acquisition of the backward curve, i. e. on the time spent to complete the curve : ~5 min (the typical time; green) or ~45 min (orange).
(B) "Positive" shifts at half height are plotted *versus* numbers of *total* (left panel) or *regularly-spaced* nucleosomes (right panel; see Equations (1) and (2)) for 28 "190-" (blue) and "208-bp" (red) fibers. The slope of the linear fit is 1.3 ±0.1 turn per regular nucleosome. Black arrowheads correspond to fibers 1-4 in (A).

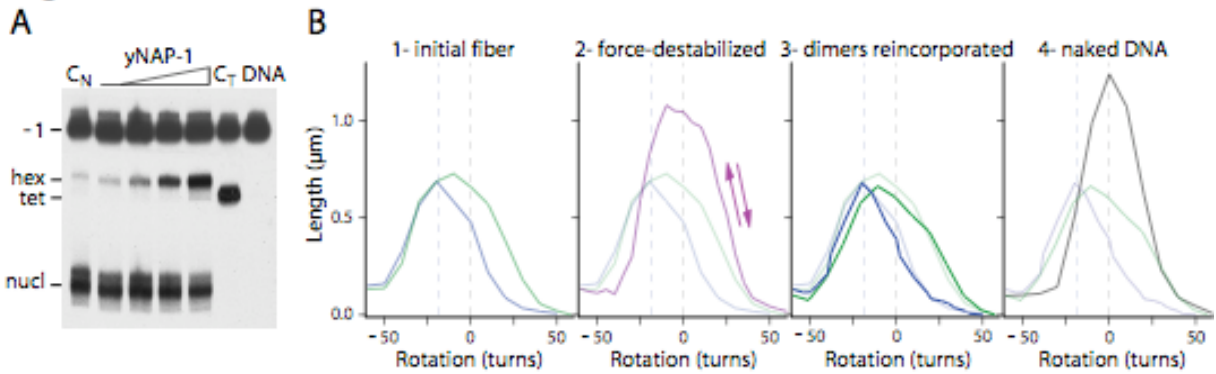

Figure 3. H2A-H2B release with yNAP-1, and reincorporation

(A) Mononucleosomes were reconstituted using the salt-jump method with pUC18 plasmid DNA as a carrier on a $\Delta Lk$ ~-1 topoisomer of a DNA minicircle formed by circularization of a $^{32}$P-end-labelled 357 bp fragment containing the 5S 208-bp repeat unit of the present arrays (Duband-Goulet et al., 1992). Reconstitution products were incubated with yeast NAP-1 at molar ratios of 0, 2.5, 7 and 20 dimers per histone octamer for 1h at 37°C in $B_0$ plus 100 mM NaCl, and electrophoresed in a 4 % (w/v) polyacrylamide gel in TE, along with starting topoisomer (DNA) and mononucleosomes (nucl; $C_N$), and control (H3-H4)$_2$ tetrasomes (tet; $C_T$). The first dimer is removed by the heparin treatment, giving rise to hexasomes (hex), but not the second. Dimer removal is actually facilitated by the open, uncrossed state of nucleosomes on topoisomer -1, as compared to nucleosomes in the negatively-crossed state on topoisomer -2 which show no release upon incubation with 20 NAP-1 dimers per octamer (Conde e Silva et al., 2007).

(B) Torsional behavior of a "190-bp" fiber at 0.2 pN through the successive steps of the assay. (1): Fiber initial response in $B_0$. (2): The force was increased to 3.5 pN at the center of rotation in $B_0$ plus 50 mM NaCl and 250 nM yNAP-1 dimers, and maintained until the first ~25 nm steps signaling the removal of tetramers (Brower-Toland et al., 2002; Claudet et al., 2005) were observed (data not shown). The force was then decreased to its original value and the flow-cell rinsed with $B_0$. Following excursion at high torsion, the hysteresis is no longer observed, i. e. backward and onward curves coincide and show an extension of the fiber (purple). (3): The fiber initial length and hysteretic response are rescued when the flow cell is filled with 40 nM H2A-H2B chaperoned with 40 nM yNAP-1 dimers in $B_0$ plus 50 mM NaCl, and flushed with $B_0$ after 5 min (bold curves). (4): Corresponding naked DNA (black).

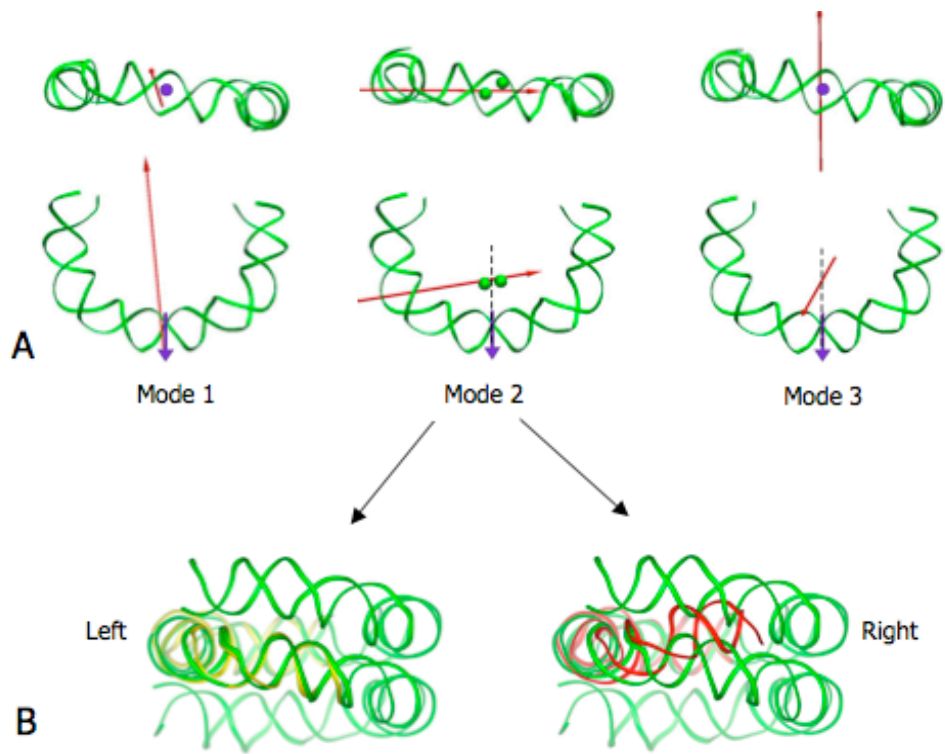

Figure 4. Normal Mode analysis of the tetrasome
(A) The axes of the rotation components of the three main vibrational modes are shown for each mode, superimposed on the tetrasome DNA superhelix viewed along the dyad (blue dots and blue arrows) or the superhelix axis. The axis of mode 1 runs close to the dyad, and the axis of mode 3 is approximately parallel to the superhelix axis. Mode 2 axis is approximately perpendicular to both dyad and superhelix axes. All three axes traverse the cysteines 110 (green balls in middle superhelix).
(B) The tetrasome was perturbed along the direction of mode 2 in (A) toward right-handedness (Right), and let relax without constraint until its energy reaches a local minimum. The resulting tetrasome DNA superhelix (red) is shown superimposed onto one side of the nucleosomal superhelix (green) viewed perpendicular to both dyad and superhelix axes. A perturbation along mode 2 toward a more left-handed superhelix (Left) does not lead to a local energy minimum, and the tetrasome goes back to its initial conformation (yellow).

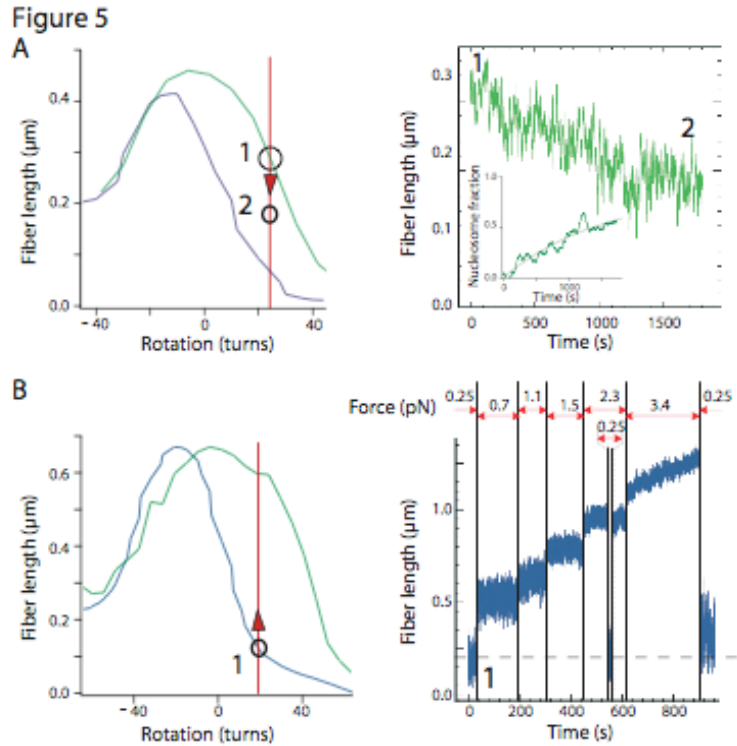

Figure 5. The transition in real time: lower salt

(A) Left: Torsional response of a "208-bp" fiber in $B_0$ at 0.4 pN. Right: The "all-altered-nucleosome" fiber at +25 turns (point 1 on the red vertical line) was allowed to relax at the same force in $B_0$ down to point 2 (green recording). Inset: proportion of nucleosomes during the relaxation time course measured as shown in Supplemental Figure S2 available with this article online, and fit with a monomolecular reaction scheme (Equation 5 in Experimental Procedures; smooth curve).

(B) Left: Same as (A) at 0.25 pN. Right: The force applied on the "all-nucleosome" fiber at +18 turns (point 1 on the red vertical line in (A)) is increased step-wise (blue recordings in right panel). A relaxation to a more extended fiber is observed when the force is raised to 3.4 pN. Upon decrease of the force back to 0.25 pN, a residual extension of ~150 nm is observed relative to point 1, corresponding to ~30% of the particles being in the altered state (measured as in (A)).

Figure 6

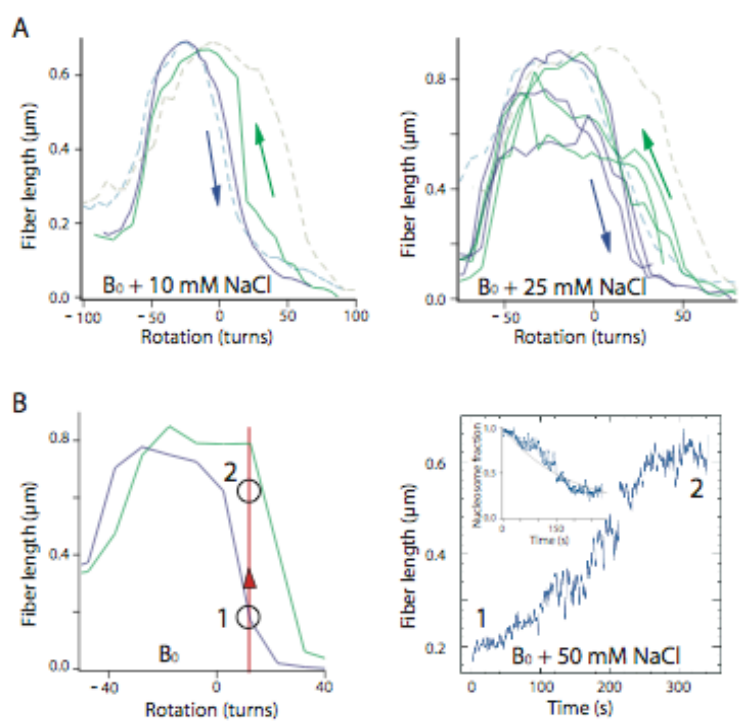

Figure 6. The transition in real time: higher salt
(A) Torsional responses of two fibers at 0.3 pN in $B_0$ (dashed) and $B_0$ + 10 mM NaCl or 25 mM NaCl (continuous). Three successive torsional cycles are shown in right panel.
(B) Left: Same as Figure 5A. Right: The flow-cell was flushed with $B_0$ + 50 mM NaCl, while the fiber was kept at a negative torsional constraint and at ~0.1 pN. A torsion of +12 turns was applied in ~5 sec (reaching point 1 on the red vertical line), and the length of the fiber was recorded in real time at 0.4 pN down to point 2 (blue tracing). Inset: fraction of nucleosomes measured as in Figure 5A from the torsional response in $B_O$ (left), and fit (smooth curve).

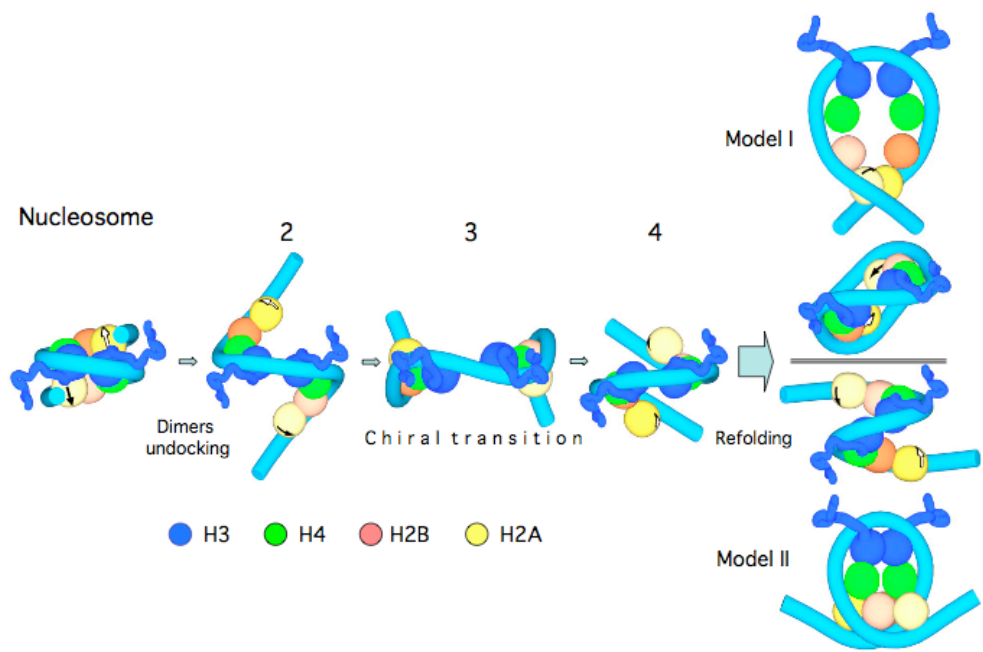

Figure 7. Putative scenario for the transition
Scheme of main steps of the transition for a nucleosome viewed as a sequential arrangement of the histones on the DNA. H2As and H2Bs in nucleosome upper and lower faces were differentiated by light and dark colors, and for H2As also by arrows. The two distal 10-bp DNAs are straight in scheme 2 and further as a result of the breaking of the H3 αN/entry-exit DNA binding sites. Two alternative routes for the refolding of the dimers are shown beyond scheme 4. In model I, entering and exiting DNAs with the dimers bound tend to wind around each other. In model II, the DNAs somehow try to continue the tetrasome right-handed superhelix. In both cases, the particle is expected to remain relatively open (see Discussion). The DNA diameter was made smaller compared to the overall particle dimensions in order to better show the histones.

**Supplemental Data**
Supplemental Data include a discussion and two figures, and can be found with this article online at http://www.molecule.org/cgi/content/full/27/1/135/DC1/

**Acknowledgments**
AB, GW, NCeS, HW and JM thank the French ministry of research, and NCeS the Foundation for medical research (FRM) and the Association for cancer research (ARC), for fellowships. AP is grateful to J. Parello for discussions and to S. Leuba (University of Pittsburg, USA) for the gift of yNAP-1, and CL and ELC to A. Justome and S. Baconnais for technical assistance. This work was supported by grants from the CNRS/MENRT programs "DRAB" and Institut Curie cooperative program "Physics of the cell" (JLV lab) and by CNRS (AP and JMV labs).

**References**

Alilat, M., Sivolob, A., Révet, B. and Prunell, A. (1999). Nucleosome dynamics. IV. Protein and DNA contributions in the chiral transition of the tetrasome, the histone (H3-H4)2 tetramer-DNA particle. J. Mol. Biol. *291*, 815-841.

Baer, B. W. and Rhodes, D. Eukaryotic RNA polymerase II binds to nucleosome cores from transcribed genes. Nature *301*, 482-488 (1983).

Bahar, I. and Rader, A. J. (2005). Coarse-grained normal mode analysis in structural biology. Curr. Opin. Struct. Biol. *15*, 586-592.

Bancaud, A., Conde e Silva, N., Barbi, M., Wagner, G., Allemand, J.-F., Mozziconacci, J., Lavelle, C., Croquette, V., Victor, J.-M., Prunell, A. and Viovy, J.-L. (2006). Structural plasticity of single chromatin fibers revealed by torsional manipulation. Nat. Struct. Mol. Biol. *13*, 444-450.

Baneres, J. L., Parello, J., Zaccai, J. and Svergun, D. (2001). A neutron scattering study of the histone sub-assemblies within the nucleosome protein core. In ILL Millenium Symposium & European User Meeting, A.J. Dianoux, ed. (I.L.L. Grenoble, France), pp.55-57.

Bazett-Jones, D. P., Mendez, E., Czarnota, G. J., Ottensmeyer, F. P. and Allfrey, V. G. (1996). Visualization and analysis of unfolded nucleosomes associated with transcribing chromatin. Nucl. Acids Res. *24*, 321-329.

Benedict, R. C., Moudrianakis, E. N. and Ackers, G. K. Interactions of the nucleosomal core histones: A calorimetric study of octamer assembly. Biochemistry *23*, 1214-1218 (1984).

Bertin, A., Leforestier, A., Durand, D. and Livolant, F. (2004). Role of histone tails in the conformation and interactions of nucleosome core particles. Biochemistry *43*, 4773-4780.

Bondarenko, V. A., Steele, L. M., Ujvari, A., Gaykalova, D. A., Kulaeva, O. I., Polikanov, Y. S., Luse, D. S. and Studitsky, V. M. (2006). Nucleosomes can form a polar barrier to transcript elongation by RNA polymerase II. Mol. Cell *24*, 469-479.

Brooks, B., Bruccoleri, R., Olafson, B., States, D., Swaminathan, S., Karplus, M. (1983). CHARMM: a program for macromolecular energy, minimization and molecular dynamics calculations. J. Comp. Chem. *4*, 187-217.

Brower-Toland, B. D., Smith, C. L., Yeh, R. C., Lis, J. T., Peterson, C. L. and Wang, M. D. (2002). Mechanical disruption of individual nucleosomes reveals a reversible multistage release of DNA. Proc. Natl. Acad. Sci. USA *99*, 1960-1965.

Claudet, C., Angelov, D., Bouvet, P., Dimitrov, S. and Bednar, J. (2005). Histone octamer instability under single molecule experiments conditions. J. Biol. Chem. *280*, 19958-19965.

Conde e Silva, N., Black, B. E., Sivolob, A., Filipski, J., Cleveland, D. W. and Prunell, A. (2007). CENP-A-containing nucleosomes: easier disassembly *versus* exclusive centromeric localization. J. Mol. Biol. *370*, 555-573.

Duband-Goulet, I., Carot, V., Ulyanov, A. V., Douc-Rasy, S. and Prunell, A. (1992). Chromatin reconstitution on small DNA rings. IV. DNA supercoiling and nucleosome sequence preference. J. Mol. Biol. *224*, 981-1001.

Epshtein, V. and Nudler, E. (2003). Cooperation between RNA polymerase molecules in transcription elongation. Science *300*, 801-805.

Hamiche, A., Carot, V., Alilat, M., De Lucia, F., O'Donohue, M. F., Révet, B. and Prunell A. (1996). Interaction of the histone (H3-H4)2 tetramer of the nucleosome with positively supercoiled DNA minicircles: Potential flipping of the protein from a left- to a right-handed superhelical form. Proc. Natl. Acad. Sci. USA *93*, 7588-7593.

Harada, Y., Ohara, O., Takatsuki, A., Itoh, H., Shimamoto, N. and Kinosita, K. Jr. (2001). Direct observation of DNA rotation during transcription by Escherichia coli RNA polymerase. Nature *409*, 113-115.

Huynh, V. A., Robinson, P. J. and Rhodes, D (2005). A method for the in vitro reconstitution of a defined "30 nm" chromatin fiber containing stoichiometric amounts of the linker histone. J. Mol. Biol. *345*, 957-968.

Ishimi, Y. and Kikuchi, A. (1991). Identification and molecular cloning of yeast homolog of nucleosome assembly protein I which facilitates nucleosome assembly in vitro. J. Biol. Chem. *266*, 7025-7029.

Ito, T., Ikehara, T., Nakagawa, T., Kraus, W. L. and Muramatsu, M. (2000). p300-mediated acetylation facilitates the transfer of histone H2A-H2B dimers from nucleosomes to a histone chaperone. Genes Dev. *14*, 1899-1907.

Jackson V. (1993). Influence of positive stress on nucleosome assembly. Biochemistry, *32*, 5901-5912.

Kepert, J. F., Mazurkiewicz, J., Heuvelman, G. L., Toth, K. F. and Rippe, K. (2005). NAP1 modulates binding of linker histone H1 to chromatin and induces an extended chromatin fiber conformation. J. Biol. Chem. *280*, 34063-34072.

Kimura, H. and Cook, P. R. (2001). Kinetics of core histones in living human cells: little exchange of H3 and H4 and some rapid exchange of H2B. J. Cell Biol. *153*, 1341-1353.

Kireeva, M. L., Walter, W., Tchernajenko, V., Bondarenko, V., Kashlev, M. and Studitsky, V. M. (2002). Nucleosome remodeling induced by RNA polymerase II: loss of the H2A/H2B dimer during transcription. Mol. Cell *9*, 541-542.

Korolev, N., Lyubartsev, A. P. and Nordenskiold, L. (2006). Computer modelling demonstrates that electrostatic attraction of nucleosomal DNA is mediated by histone tails. Biophys. J. *90*, 4305-4316.

Lee, M. S. and Garrard, W. T. (1991). Positive DNA supercoiling generates a chromatin conformation characteristic of highly active genes. Proc. Natl. Acad. Sci. USA *88*, 9675-9679.

Li, G., Levitus, M., Bustamante, C. and Widom, J. (2005) Rapid spontaneous accessibility of nucleosomal DNA. Nat. Struct. Mol. Biol. *12*, 46-53.

Liu, L. F. and Wang, J. C. (1987). Supercoiling of the DNA template during transcription. Proc. Natl. Acad. Sci. USA *84*, 7024-7027.

Ljungman, M. and Hanawalt, P. C. (1992). Localized torsional tension in the DNA of human cells. Proc. Natl. Acad. Sci. USA *89*, 6055-6059.

Louters, L. and Chalkley, R. (1985). Exchange of histones H1, H2A, and H2B in vivo. Biochemistry *24*, 3080-3085.

Luger, K., Mader, A. W., Richmond, R. K., Sargent, D. F. and Richmond, T. J. (1997). Crystal structure of the nucleosome core particle at 2.8 A resolution. Nature *389*, 251-260.

McKerell, A. D. Jr., Brooks, B., Brooks, C. L. III, Nilsson, L., Roux, B., Won, Y. and. Karplus, M. (1998a). CHARMM: The energy function and its parameterization with an overview of the program. In The encyclopedia of computational chemistry, P. v. R. Schleyer et al. eds. (John Wiley and Sons, Chichester), pp. 271-277.

MacKerell, A. D. Jr., Bashford, D., Bellott, M., Dunbrack, R. L. Jr., Evanseck, J. D., Field, M. J., Fischer, S., Gao, J., Guo, H., Ha, S. et al. (1998b). All-atom empirical potential for molecular modelling and dynamics studies of proteins. J. Phys. Chem. B *102*, 3586-3616.

Mihardja, S., Spakowitz, A. J., Zhang, Y. and Bustamante, C. (2006). Effect of force on mononucleosomal dynamics. Proc. Natl. Acad. Sci. U S A. *103*, 15871-15876.

Mouawad, L. and Perahia, D. (1993). Diagonalization in a mixed basis: a method to compute low-frequency normal modes for large macromolecules. Biopolymers *33*, 599-611.

Mouawad, L. and Perahia, D. (2006). All-atom normal mode calculations of large molecular systems using iterative methods. In Normal mode analysis: Theory and applications to biological and chemical systems, Q. Cui and I. Bahar, eds (C&H/CRC, Mathematical & Computational Biology Series, CRC Press), pp.17-39.

Neukirch, S., van der Heijden, G.H.M. and Thompson, J.M.T. (2002). Writhing instabilities of twisted rods: from infinite to finite lengths. J. Mech. Phys. Solids *50*, 1175-1191.

Pfaffle, P., Gerlach, V., Bunzel, L. and Jackson, V. (1990). In vitro evidence that transcription-induced stress causes nucleosome dissolution and regeneration. J. Biol. Chem. *265*, 16830-16840.

Protacio, R. U. and Widom, J. (1996). Nucleosome transcription studied in a real-time synchronous system: test of the lexosome model and direct measurement of effects due to histone octamer. J. Mol. Biol. *256*, 458-472.

Reinberg, D. and Sims, R. J. 3rd. (2006). de FACTo nucleosome dynamics. J. Biol. Chem. *281*, 23297-23301.

Ruiz-Carrillo, A. and Jorcano, J. L. (1978). Nucleohistone assembly: sequential binding of histone H3-H4 tetramer and histone H2A-H2B dimer to DNA. Cold Spring Harb. Symp. Quant. Biol. *42*, 165-170.

Salceda, J., Fernandez, X. and Roca, J. (2006). Topoisomerase II, not topoisomerase I, is the proficient relaxase of nucleosomal DNA. EMBO J. *25*, 2575-2583.

Schwager, S., Retief, J. D., de Groot, P. and von Holt, C. (1985). Rapide exchange of histones H2A and H2B in sea urchin embryo chromatin. FEBS Lett. *189*, 305-309.

Sivolob, A. and Prunell, A. (2004). Nucleosome conformational flexibility and implications for chromatin dynamics. Philos. Transact. A Math. Phys. Eng. Sci. *362*, 1519-1547.

Sivolob, A., De Lucia, F., Alilat, M. and Prunell, A. (2000). Nucleosome dynamics. VI. Histone tail regulation of tetrasome chiral transition. A relaxation study of tetrasomes on DNA minicircles. J. Mol. Biol. *295*, 55-69.

Strick, T. R., Allemand, J. F., Bensimon, D., Bensimon, A. and Croquette, V. (1996). The elasticity of a single supercoiled DNA molecule. Science *271*, 1835-1837.

Studitsky, V. M., Kassavetis, G. A., Geiduschek, E. P. and Felsenfeld, G. (1997). Mechanism of transcription through the nucleosome by eukaryotic RNA polymerase. Science, *278*, 1960-1963.

Tomschik, M., Zheng, H., van Holde, K., Zlatanova, J. and Leuba, S. H. (2005). Fast, long-range, reversible conformational fluctuations in nucleosomes revealed by single-pair fluorescence resonance energy transfer. Proc. Natl Acad. Sci. USA, *102*, 3278-3283.

van Holde, K. E. (1988). *Chromatin* (Springer-Verlag, New York).

Wilhelm, F. X., Wilhelm, M. L., Erard, M. and Daune, M. P. (1978). Reconstitution of chromatin: assembly of the nucleosome. Nucleic Acids Res. *5*, 505-521.

Wilson, E. B., Decius, J.C. and Cross, P.C. (1955). Molecular vibrations: The theory of infrared and Raman vibrational spectroscopy (Dover Publications Inc., N.Y.).

Wriggers, W. and Schulten, K. (1997). Protein domain movements: Detection of rigid domains and visualization of effective rotations in comparisons of atomic coordinates. Proteins *29*, 1-14.

Yodh, J.G., Woodbury, N., Shlyakhtenko, L.S., Lyubchenko, Y.L. and Lohr, D. (2002). Mapping nucleosome locations on the 208-12 by AFM provides clear evidence for cooperativity in array occupation. Biochemistry *41*, 3565-3574.